\documentclass[conference]{IEEEtran}
\IEEEoverridecommandlockouts
\usepackage{cite}
\usepackage{amsmath,amssymb,amsfonts}
\usepackage{algorithmic}
\usepackage{graphicx}
\usepackage{textcomp}
\usepackage{xcolor}
\usepackage{multirow}
\usepackage{subcaption}
\usepackage{adjustbox}

\usepackage{tabularx}
\newcolumntype{L}{>{\centering\arraybackslash}X}

\usepackage{todonotes}
\newcommand{\bfA}{{\bf A}}

\newcommand{\bfI}{{\bf I}}
\newcommand{\bfJ}{{\bf J}}

\newcommand{\bfV}{{\bf V}}
\newcommand{\bfW}{{\bf W}}
\newcommand{\bfX}{{\bf X}}
\newcommand{\bfY}{{\bf Y}}

\newcommand{\bfx}{{\bf x}}

\newcommand{\ie}{i.e., }
\newcommand{\eg}{e.g., }
\usepackage{mathtools}

\def\BibTeX{{\rm B\kern-.05em{\sc i\kern-.025em b}\kern-.08em
    T\kern-.1667em\lower.7ex\hbox{E}\kern-.125emX}}
\begin{document}

\title{Non-Dissipative Graph Propagation for Non-Local Community Detection}

 \author{\IEEEauthorblockN{William Leeney}
 \IEEEauthorblockA{\textit{School of Engineering Mathematics and Technology} \\
 \textit{University of Bristol}\\
 Bristol, UK \\
 will.leeney@bristol.ac.uk}
 \and
 \IEEEauthorblockN{Alessio Gravina}
 \IEEEauthorblockA{\textit{Dept. of Computer Science} \\
 \textit{University of Pisa}\\
 Pisa, Italy \\
 alessio.gravina@di.unipi.it}
 \and
 \IEEEauthorblockN{Davide Bacciu}
 \IEEEauthorblockA{\textit{Dept. of Computer Science} \\
 \textit{University of Pisa}\\
 Pisa, Italy \\
 davide.bacciu@unipi.it}
 }

\maketitle

\begin{abstract}
Community detection in graphs aims to cluster nodes into meaningful groups, a task particularly challenging in heterophilic graphs, where nodes sharing similarities and membership to the same community are typically distantly connected. This is particularly evident when this task is tackled by graph neural networks, since they rely on an inherently local message passing scheme to learn the node representations that serve to cluster nodes into communities.  In this work, we argue that the ability to  propagate long-range information during message passing is key to effectively perform community detection in heterophilic graphs. To this end, we introduce the Unsupervised Antisymmetric Graph Neural Network (uAGNN), a novel unsupervised community detection approach leveraging non-dissipative dynamical systems to ensure stability and to propagate long-range information effectively. 
By employing antisymmetric weight matrices, uAGNN captures both local and global graph structures, overcoming the limitations posed by heterophilic scenarios. Extensive experiments across ten datasets demonstrate uAGNN's superior performance in high  and medium heterophilic settings, where traditional methods fail to exploit long-range dependencies. These results highlight uAGNN's potential as a powerful tool for unsupervised community detection in diverse graph environments.


\end{abstract}


\section{Introduction}

Graphs are expressive modeling tools for representing and understanding high-dimensional relationships in data. For example, social networks allow us to understand the spread of misinformation \cite{benamira2019semi} and in biology, graphs can help us learn how drugs interact with the body \cite{li2022geomgcl, gravina_covid, gravina_schizophrenia}. Finding communities of similar nodes in graphs, sometimes known as node clustering, is a challenging task with high practical implications as it allows to identify patterns and regularities in highly complex relational data. For instance,  in protein-protein interaction networks the identification of similar proteins can guide us in the development of new treatments to disease \cite{manipur2021community}; in online social networks, the identification of communities of users serves to promote targeted interactions \cite{bedi2016community}. 

In the real-world, it is desirable to apply community detection algorithms to find communities where the ground-truth partitioning is unknown. Recently, graph neural networks (GNNs)~\cite{BACCIU2020203,gravina_dynamic_survey} have been proposed as an effective tool to learn unsupervised representations of graphs for the purpose of solving community detection tasks \cite{wang2019attributed, tsitsulin2023graph, qiu2022VGAER, kipf2016variational, velickovic2019deep, hassani2020contrastive, zhu2020deep, thakoor2021bootstrapgraph, liu2022towards}. These approaches, however, have been developed under the assumption that nodes in the same community tend to be locally connected, which may not be true for all graphs. In online purchasing networks, fraudsters typically build connections with customers rather than other fraudsters \cite{pandit2007netprobe}. Protein structures are naturally composed of different types of amino acids that are locally connected \cite{zhu2020beyond}. In this work, we argue that whenever faced with community detection on graphs where similar nodes are non-locally adjacent, we need to resort to GNNs that are effective in propagating and exploiting long-range node relationships when building their representations.  
This is because communities cannot be assumed to be clusters of locally connected nodes. 
This particular setting resembles that of learning graph representations where adjacent nodes belong to different classes, commonly referred to as heterophilic (or low-homophilic) setting. Such a scenario, and its impact on the quality of the representations learned by different GNN models, has been widely investigated for supervised tasks \cite{zheng2022graph}, but has not yet been evaluated in unsupervised community detection problems, which is one of the key contributions of our work.

Based on these considerations, we propose \textit{Unsupervised Antisymmetric Graph Neural Network} (uAGNN), a novel unsupervised GNN-based approach for community detection that leverages non-local graph information for 
reconstructing both adjacency and feature matrices, enabling robust representation learning without supervision. Leveraging the connection with non-dissipative dynamical systems, our uAGNN  employs antisymmetric weight matrices to ensure stability and non-dissipative behaviors, enabling effective information propagation over long radii in the graph. As we will show through our experimental analysis, this capability is fundamental for identifying communities that are weakly connected via adjacency information.

The key contributions of this work are:
\begin{itemize}
    \item We introduce uAGNN, a novel GNN for community detection that is able to leverage and preserve long-range dependencies in the information flow, facilitating effective community detection in complex graph structures. 
    \item We provide empirical evidence demonstrating the critical importance of long-range information propagation in scenarios where nodes belonging to the same community are connected through longer and sparser paths, \ie heterophilic settings.
    \item We perform extensive experimental evaluations across diverse datasets to demonstrate the benefits of our method, showing that uAGNN achieves superior performance, particularly as the level of heterophily increases. 
\end{itemize}
To the best of our knowledge, this is the first work exploring the role of heterophily in unsupervised community detection and highlighting the relevance of long-range information propagation also for unsupervised learning tasks.

\section{Preliminaries}

In this work, we consider a graph $\mathcal{G} = (\mathcal{V}, \mathcal{E}, \bfX)$ as a system of interacting entities, referred to as nodes. Specifically, a graph is defined by a set $\mathcal{V}$ of $n$ nodes and a set of edges $\mathcal{E}\subseteq\mathcal{V}\times\mathcal{V}$, which explicitly defines the connections between nodes. The relational information between nodes can also be represented by the adjacency matrix $\bfA \in\{0,1\}^{n \times n}$, where each entry $\bfA_{u,v}$ is defined as $1$ if $(u,v) \in \mathcal{E}$, and $0$ otherwise. A graph is undirected if $\bfA$ is symmetric, \ie $\bfA_{u,v}=\bfA_{v,u}$.  Each node in a graph has a set of attributes (also called features), $\bfx_u\in\mathbb{R}^{d}$ with $d$ the number of node features, which represent the node state. All the node states are stored in the matrix $\bfX\in\mathbb{R}^{n \times d}$. We define the neighborhood of a node $u$, $\mathcal{N}_u$, all the nodes that are directly connected to $u$, \ie $\mathcal{N}_u = \{v \;|\; \bfA_{vu} = 1 \}$. Lastly, we note that a graph is said to exhibit a high level of homophily if nodes within the same neighborhood are likely to belong to the same class or share similar features. Conversely, the graph is said to exhibit a high level of heterophily.

The main objective of this work is to perform community detection on undirected graphs. In other words, the goal is to partition the graph $\mathcal{G}$ into $k$ clusters such that nodes in each partition, or cluster, exhibit similar structure and states, relying only on the graph information and the predefined number of clusters to partition the graph into. In the following, we assume the task of hard clustering. Specifically, each node is to be associated with a single community, such that $\bfY \in \{1,2\dots,k\}^{n}$.



\section{The Method}
As emerged in recent literature \cite{alon2021oversquashing, LRGB} the sole exploitation of local interaction between nodes is not enough for learning effective representations. In this scenario, several methods have been proposed to overcome this limitation \cite{topping2022understanding, chamberlain2021grand, dwivedi2021generalization, drew, rampasek2022GPS, maskey2023fractional}. Specifically, the family of models that couples stable and non-dissipative Ordinary Differential Equations (ODEs) with Graph Neural Networks (GNNs) \cite{gravina_adgn, gravina_swan, gravina_phdgn} has emerged as one of the most effective for exploiting and propagating non-local node information. This effectiveness stems from their strong theoretical foundation, flexibility, and computational efficiency.

Building on this line of work, we now introduce our framework, which leverages the non-dissipative GNN presented in \cite{gravina_adgn} as its backbone. We start by presenting the connection between GNNs,  information diffusion in graphs, and dynamical systems, \ie systems whose states are described by differential equations. 
\\

\begin{figure*}[h!]
    \includegraphics[width=0.98\textwidth]{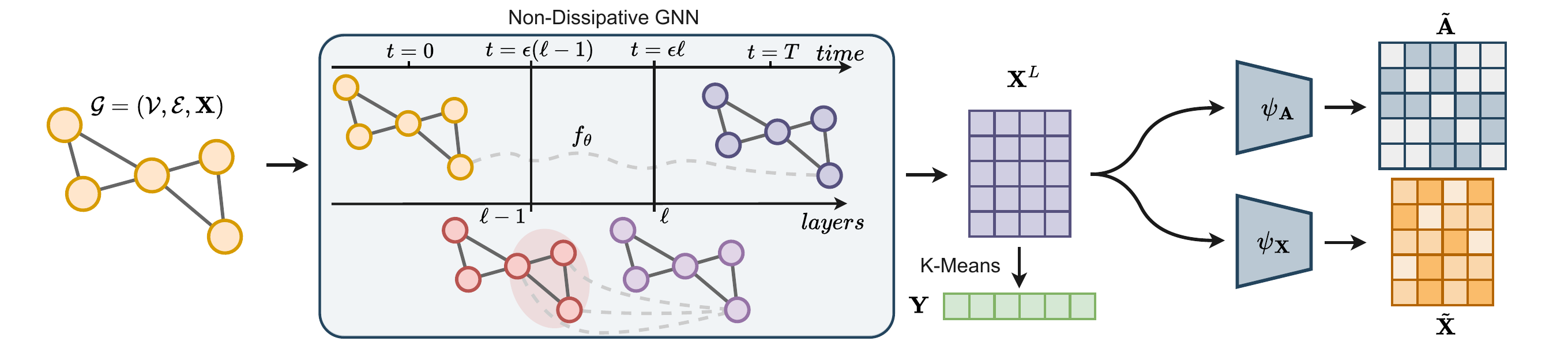}
    \caption{The architecture of the Unsupervised Antisymmetric Graph Neural Network (uAGNN). Given an input graph $\mathcal{G}$, our uAGNN encodes first the attributed graph into a new representation space $\bfX^L$ using a non-dissipative GNN, which effectively captures both short- and long-range node information. Subsequently, the new node representations $\bfX^L$ are utilized for two purposes: (i) to decode the reconstructed adjacency $\tilde{\bfA}$ and input feature space $\tilde{\bfX}$ via two separate linear transformations (\ie $\psi_\bfA$ and $\psi_\bfX$), and (ii) to obtain the node cluster assignments $\bfY$ using the K-Means algorithm.}
\label{fig:architecture}
\end{figure*}

\textbf{Modeling GNNs as Dynamical Systems.} Without loss of generality, a dynamical system associated to information diffusion can be described using the following node-wise ODE:
\begin{equation}\label{eq:general_ode}
    \begin{dcases}
    \frac{d\bfx_u(t)}{dt} = f_\theta(t, \bfx_u(t), \{\bfx_v(t)\}_{v\in\mathcal{N}_u}), &  t\in[0,T]\\
    \bfx_u(0) = \bfx_u^0
    \end{dcases}
\end{equation}
where $f_\theta: \mathbb{R}^d\rightarrow \mathbb{R}^d$ is a function parametrized by the weights $\theta$ that describes the dynamics of node states, subject to the initial condition $\bfx_u^0\in \mathbb{R}^d$. Consequently, the ODE in Equation~\ref{eq:general_ode} functions as a continuous system for processing information across the graph. Beginning with the initial input configuration $\bfx_u(0)$, it evolves the node states to the final representation (\ie embedding) of each node, $\bfx_u(T)$.
As demonstrated in \cite{gravina_adgn}, this process aligns with standard GNNs in how it evolves node representations, which can subsequently be used as embeddings for solving downstream tasks on the graph. Specifically, these similarities arise from the use of numerical discretization methods to solve the ODE associated with the dynamical system. The numerical method computes the ODE solution iteratively over a discrete set of points in $[0, T]$ with a step size $\epsilon>0$. In this context, each step of the discretization process corresponds to one layer of the GNN, meaning that the node representation at layer $\ell$, $\bfx_u^\ell$, approximates the computation of $\bfx_u(\epsilon\ell)$ in Equation~\ref{eq:general_ode}. The middle of Figure~\ref{fig:architecture} shows this process visually.\\

\textbf{Non-Dissipation in GNNs.} We now proceed to derive the conditions under which the ODE in Equation~\ref{eq:general_ode} is constrained to a stable and non-dissipative behavior, allowing for the propagation of long-range dependencies in the information flow. 

We start by instantiating Equation~\ref{eq:general_ode} as:
\begin{equation}\label{eq:adgn_ode}
    \frac{d\bfx_u(t)}{dt} = \sigma(\bfW_t\bfx_u(t), \Phi(\{\bfx_v(t)\}_{v\in\mathcal{N}_u}))
\end{equation}
where $\sigma$ is the activation function, $\bfW_t\in\mathbb{R}^{d\times d}$ is the weight matrix, and $\Phi$ is the aggregation function that computes the representation of the neighborhood of the node $u$\footnote{We note that $\Phi$ can be any function that aggregates node and edges information. Therefore, we can leverage the aggregation function that is more suitable for the problem. As a proof of this, we the experimental section we explore two aggregation functions derived by \cite{morris2019weisfeiler, GCN}.}.

As discussed in \cite{HaberRuthotto2017, chang2018antisymmetricrnn, gravina_adgn}, a non-dissipative propagation is directly linked with the sensitivity of the solution of the ODE to its initial condition, thus the stability of the system, \ie:
\begin{equation}
    \frac{d}{dt}\left(\frac{d\bfx_u(t)}{d\bfx_u(0)}\right) = \mathbf{J}(t)\frac{d\bfx_u(t)}{d\bfx_u(0)}
\end{equation}
Such sensitivity is controlled by the eigenvalues of the Jacobian $\bfJ$ of Equation~\ref{eq:adgn_ode} (\ie $\lambda_i$), leading to three different behaviors: (i) instability, (ii) dissipativity, and (iii) non-dissipativity.

\textbf{Instability} is observed when $max_{i=1,\dots,d} Re(\lambda_i) > 0$. In this case, a small perturbation of the initial condition causes an exponential divergence in node representations.

\textbf{Dissipativity} is determined when $max_{i=1,\dots,d} Re(\lambda_i) < 0$. Only local information neighborhood information is preserved by the system, making the node representations insensitive to differences in the input graph. Therefore, the final GNN results in a lossy system that is not capable of propagating long-range information across the graph.

\textbf{Non-Dissipativity} is obtained when $Re(\lambda_i) = 0$. Here, the system is stable and the input graph information is effectively propagated through the successive transformations into the final node representations. In other words, the GNN is able to preserve long-range dependencies between nodes.

We note that by constraining $\bfW_t$ to be antisymmetric (\ie $-\bfW_t = \bfW_t^\top$) we obtain the ODE in Equation~\ref{eq:adgn_ode} is non-dissipative, since the resulting Jacobian has either purely imaginary eigenvalues, in the case the aggregation function $\Phi$ does not include $\bfx_u$ itself, or bounded eigenvalues in a small neighborhood of the imaginary axes, if the aggregation include $\bfx_u$ \cite{gravina_adgn}. 
\\

\textbf{Numerical Discretization.} Now that we have defined the conditions under which the ODE in Equation~\ref{eq:adgn_ode} is stable and non-dissipative, i.e., it can propagate long-range dependencies between nodes, as we previously observed, we note that computing the analytical solution of an ODE is usually infeasible. We employ the \textit{forward Euler's method} to discretize Equation~\ref{eq:adgn_ode} to obtain the following node state update equations for the node $u$ at layer $\ell$:
\begin{equation}\label{eq:adgn_discretized}
\begin{split}
        \bfx_u^\ell = \bfx_u^{\ell-1}+\epsilon\sigma((\bfW_t-\bfW_t^\top&-\gamma\bfI)\bfx_u^{\ell-1}\\
        &+\Phi(\{\bfx_v^{\ell-1}\}_{v\in\mathcal{N}_u})),
\end{split}
\end{equation}
where the term $-\gamma\bfI$ (with $\bfI$ the identity matrix of dimension $d\times d$) is introduced to preserve the stability of the forward Euler's method \cite{Ascher1998}. $\gamma\in\mathbb{R}$ is a hyperparameter that regulates the strength of the stability of the discretization method. We note that the computational complexity of Equation~\ref{eq:adgn_discretized} is governed by the aggregation function $\Phi$, as the additional skip connection and matrix operations have a negligible impact on the overall complexity. In particular, when using MPNN-like aggregation functions, Equation~\ref{eq:adgn_discretized} retains the complexity of an MPNN.
\\

\textbf{The Unsupervised Adaptation of Non-Dissipative GNNs.}
We now turn to preset the overall framework that adapts non-dissipative GNNs to the unsupervised setting. Unlike previous autoencoder frameworks, which focus on reconstructing either the adjacency information \cite{kipf2016variational} or the feature matrix \cite{wang2017mgae}, our approach reconstructs both simultaneously. In order to do so, we first employ the non-dissipative GNN in Equation~\ref{eq:adgn_discretized} to compute the new node representations, \ie $\bfX^L$, encoding both short and long-range information. Then, we employ two learned linear transformations to reconstruct the initial feature matrix $\bfX$ and the connectivity of the graph $\bfA$. More formally, denoting with $\psi_\bfX: \mathbb{R}^{n\times d}\rightarrow \mathbb{R}^{n\times d}$ and $\psi_\bfA: \mathbb{R}^{n\times d} \rightarrow \mathbb{R}^{n\times n}$ the two transformations, we compute 
\begin{equation}
\tilde{\bfX} = \sigma(\psi_{\bfX}(\bfX^L)),\quad \tilde{\bfA} = \sigma(\psi_{\bfA}(\bfX^L)).
\end{equation}
To enforce the model to minimize the reconstruction error of the original node representations and adjacency matrix, we train the non-dissipative GNN as well as the linear transformations in an end-to-end fashion according to the following loss function, which combines the mean square error (MSE) between the respective original matrices and the reconstructions:
\begin{equation}
        \mathcal{L} = \text{MSE}\big(\bfA, \tilde{\bfA}\big) + \text{MSE}\big(\bfX, \tilde{\bfX}\big).
\end{equation}
To obtain the node cluster assignments $\bfY$, we employ the K-Means algorithm applied to the final node representation $\bfX^L$, \ie $\bfY = \text{K-Means}(\bfX^L)$. Figure \ref{fig:architecture} illustrates the flow of information of the whole architecture. 

We name the framework defined above, Unsupervised Antisymmetric Graph Neural Network (uAGNN).

\section{Related work}

\subsection{Unsupervised Attributed Graph Representation Learning}

There is a wide breadth of research on GNNs for learning representations of attributed graphs without using labels in the loss function. Given the evaluation task of community detection, the algorithms which are considered comparable in this work are those that are can perform node clustering via an embedded space. In addition to algorithms explicitly designed for community detection, we also consider methods that learn unsupervised representations of data as there is previous research that applies vector-based clustering algorithms to the embeddings \cite{mcconville2021n2d,leeney2023uncertainty,leeney2024unsupervised,leeney2024investigation}.

Deep Attentional Embedded Graph Clustering (DAEGC) uses a k-means target to self-supervise the clustering module to iteratively refine the clustering of node embeddings \cite{wang2019attributed}. This links the embedding space and the graph clustering module for mutual benefit. Deep Modularity Networks (DMoN) maximize a modularity based clustering objective to optimize cluster assignments \cite{tsitsulin2023graph}. Variational Graph AutoEncoder Reconstruction (VGAER) \cite{qiu2022VGAER} learns to embed a graph by recovering a modularity matrix by using an inner product decoder from the encoding of a VGAE \cite{kipf2016variational}. Deep Graph Infomax (DGI) maximizes mutual information between mutual information of high-level summaries of graph embedding representations using by contrastive learning \cite{velickovic2019deep}. Contrastive Multi-View Representation Learning on Graphs (MVGRL) argues that the best employment of contrastive methods for graphs is achieved by contrasting encodings' from first-order neighbours and a general graph diffusion \cite{hassani2020contrastive}. GRAph Contrastive rEpresentation learning (GRACE) generates a corrupted view of the entire graph by removing edges and learns node representations by maximizing agreement across two views \cite{zhu2020deep}. Bootstrapped Graph Latents (BGRL) \cite{thakoor2021bootstrapgraph} uses a self-supervised bootstrap procedure by maintaining two graph encoders; the online one learns to predict the representations of the target encoder, which in itself is updated by an exponential moving average of the online encoder. Towards Unsupervised Deep Graph Structure Learning (SUBLIME) \cite{liu2022towards} employs a contrastive learning scheme for representation learning as well as using the bootstrapping principle to learn a graph structure from the anchor graph. 

\subsection{GNNs based on Dynamical Systems}
Recent advancements in representation learning have introduced architectures that bridge the gap between neural networks and dynamical systems. Notably, approaches such as 
GODE \cite{zhuang2020ordinary}, GRAND \cite{GRAND}, and PDE-GCN\textsubscript{D} \cite{pdegcn} 
interpret GNN layers as discretization steps of differential equations, \eg utilizing the heat equation. Other architectures, like PDE-GCN\textsubscript{M} \cite{pdegcn} and GraphCON \cite{graphcon}, incorporate a mix of diffusion and oscillatory processes to preserve feature energy. The extension to the temporal domain as been studied in works such as \cite{gravina_tgode, gravina_ctan}.

In addition, recent studies have introduced mechanisms such as anti-symmetry \cite{gravina_adgn, gravina_randomized_adgn, gravina_swan}, reaction-diffusion-based dynamics \cite{
choi2022gread}, port-Hamiltonian dynamics \cite{gravina_phdgn}, and advection-reaction-diffusion processes \cite{eliasof2023adr}. However, most of these approaches remain rooted in diffusion-based paradigms, which are inherently limited in effectively modeling long-range interactions in graphs.

\section{Experiments}
In this section, we discuss the empirical assessment of our uAGNN. 
The objective of our experiments is to show that where communities exist that are characterized by long range graph connections, uAGNN achieves better community detection performance than related approaches from literature. 
Specifically, we compare the performance of our uAGNN with respect to DAEGC \cite{wang2019attributed},   DGI\cite{velickovic2019deep},   DMoN\cite{tsitsulin2023graph}, GRACE\cite{zhu2020deep}, MVGRL\cite{hassani2020contrastive}, SUBLIME\cite{liu2022towards}, BGRL\cite{thakoor2021bootstrapgraph},  VGAER \cite{qiu2022VGAER}.
We evaluate performance across a variety of datasets exhibiting different levels of homophily, defined as the fraction of edges in a graph that connect nodes sharing the same community label, as described in \cite{zhu2020beyond} and formalized as
\begin{equation}
    \mathcal{H} = \frac{|\{ (v, u) : (v, u) \in \mathcal{E} \wedge y_v = y_u \}|}{|\mathcal{E}|},
    \label{equ:homophily}
\end{equation}
where $y_v, y_u\in\bfY$.
In the experiments within, the objective is to show that uAGNN performs favorably in graphs with low homophily which indicates that communities exist with long range connections.

\begin{table}[h]
    \caption{The resources used by all methods in the investigation and hyperparameter optimization.}
    \begin{center}
    \begin{tabularx}{\linewidth}{ |c | L | }
    \hline
    \textbf{Resource} & \textbf{Associated Allocation}  \\
    \hline
    Max Hyperparameter Trials & $300$ \\
    No. K-means Inits & 20 \\
    Seeds & 42, 24, 976, 12345, 8765, 7, 856, 90, 672, 785 \\
    Training Split & $0.64$ \\ 
    Validation Split & $0.16$ \\ 
    Testing Split & $0.2$ \\ 
    Max Epochs &  $5000$ \\
    Optimiser & Adam \cite{kingma2014adam} \\
    HP Optimiser & MOTPE \cite{ozaki2022multiobjective} \\
    Learning Rate &  0.05, 0.01, 0.005, 0.001, 0.0005, 0.0001 \\
    Weight Decay & 0.05, 0.005, 0.0005, 0.0 \\
    \hline
    \end{tabularx}
    \label{tab:hpo_resources}
    \end{center}
\end{table}
\begin{table}[h]
    \caption{The attributed community detection datasets statistics. }
    \begin{center}
        \begin{tabular}{|c|c|c|c|c|c|}
        \hline
        \textbf{Dataset}&\multicolumn{5}{|c|}{\textbf{Dataset Statistics}} \\
        \cline{2-6} 
        \textbf{Name} & \textbf{\textit{Nodes}}& \textbf{\textit{Edges}}& \textbf{\textit{Feat.}} & \textbf{\textit{Classes}} & \textbf{\textit{Homophiliy}} \\
        \hline
         AMAP \cite{he2016ups} & 7650 & 238163 & 745 & 8 & 0.83 \\
         BAT \cite{deep_graph_clustering_survey} & 131 & 2077 & 81 & 4 & 0.45 \\
         CiteSeer  \cite{giles1998citeseer} & 3327 & 9104 & 3703 & 6 & 0.74 \\
         Cora \cite{mccallum2000automating} & 2708 & 10556 & 1433 & 7 & 0.81 \\
         Cornell \cite{craven1998learning} & 183 & 298 & 1703 & 5 & 0.31 \\
         DBLP \cite{tang2008arnetminer} & 4057 & 7056 & 334 & 4 & 0.8 \\
         EAT \cite{deep_graph_clustering_survey} & 399 & 11988 & 203 & 4 & 0.4 \\
         Texas \cite{craven1998learning} & 183 & 325 & 1703 & 5 & 0.11 \\
         UAT \cite{deep_graph_clustering_survey} & 1190 & 27198 & 239 & 4 & 0.7 \\
         Wisconsin \cite{craven1998learning} & 251 & 515 & 1703 & 5 & 0.2 \\
        \hline
        \end{tabular}
    \label{tab:dataset_statistics}
    \end{center}
\end{table}
\begin{table}[h!]
    \caption{The uAGNN hyperparameter (HP) space. We note that in the GNN aggregation $\bfA$ is the original adjacency matrix, $\mathbf{\hat{A}}=\bfA+\bfI$, and $\mathbf{\hat{D}}_{ii} = \sum_{j=0} \mathbf{\hat{A}}_{ij}$ is the diagonal degree matrix.}
    \begin{center}
    \begin{tabularx}{\linewidth}{ |c | L | }
    \hline
    \textbf{HP} & \textbf{Search space}  \\
    \hline
    No. Layers & 1, 2, 3, 5, 10, 20, 30 \\
    Hidden Dim. & 32, 64, 128 \\
    $\gamma$ & 0.0001, 0.001, 0.01, 0.1, 1.\\ 
    $\epsilon$ & 0.0001, 0.001, 0.01, 0.1, 1.\\ 
    GNN Agg. & $\Phi_{1}=\bfA\bfX^{\ell-1}\bfV$,\quad $\Phi_{2}=(\mathbf{\hat{D}}^{-1/2} \mathbf{\hat{A}}
\mathbf{\hat{D}}^{-1/2})\bfX^{\ell-1}\bfV$\\ 
    \hline
    \end{tabularx}
    \label{tab:uagdn_hpo}
    \end{center}
\end{table}

\subsection{Experiment Details}

To ensure a fair evaluation, we train our uAGNN and all baseline models using the same resources, as detailed in Table~\ref{tab:hpo_resources}. The evaluation is repeated on ten different seeds, and the datasets are split into training, validation, and test sets by randomly removing edges to create each partition. Specifically, the training partition, $p_{\text{train}}$, contains 64\% of the original edges, the validation partition includes $p_{\text{train}} + p_{\text{val}}$, \ie 80\% of the edges, and the test partition comprises the remaining edges that belong exclusively to the test set. 

In Table~\ref{tab:dataset_statistics} we detail the dataset used in this investigation. As metrics of performance for our investigation we employ: 
macro-F1 (F1), which measure the performance of the method with respect to the original target class of each node; Normalized Mutual Information (NMI), which measures the mutual dependence between the elements in the clusters; and Conductance, which measures the quality of a cluster by evaluating how well it is separated from the rest of the graph. All the metrics are scaled between 0 and 1.

To accurately compare each method, we perform Hyper-Parameter (HP) optimization for all models on each dataset on each metric of performance independently, employing the Multi-Objective Tree Parzen Estimator (MOTPE) algorithm \cite{ozaki2022multiobjective}, optimizing the validation score. We detail the hyperparameter space that we explore for uAGNN in Table \ref{tab:uagdn_hpo}. 

All models are trained using a 2080 Ti GPU on a server with 12 GB of RAM and a 16-core Xeon CPU. We also make the code for the whole evaluation open-source\footnote{https://github.com/willleeney/ugle/blob/main/ugle/models/uadgn.py}.

\subsection{Experiment Results}
\begin{figure}[htpb]
\begin{adjustbox}{center}
  \begin{subfigure}{1.34\columnwidth}
    \centering \includegraphics[width=0.8\linewidth]{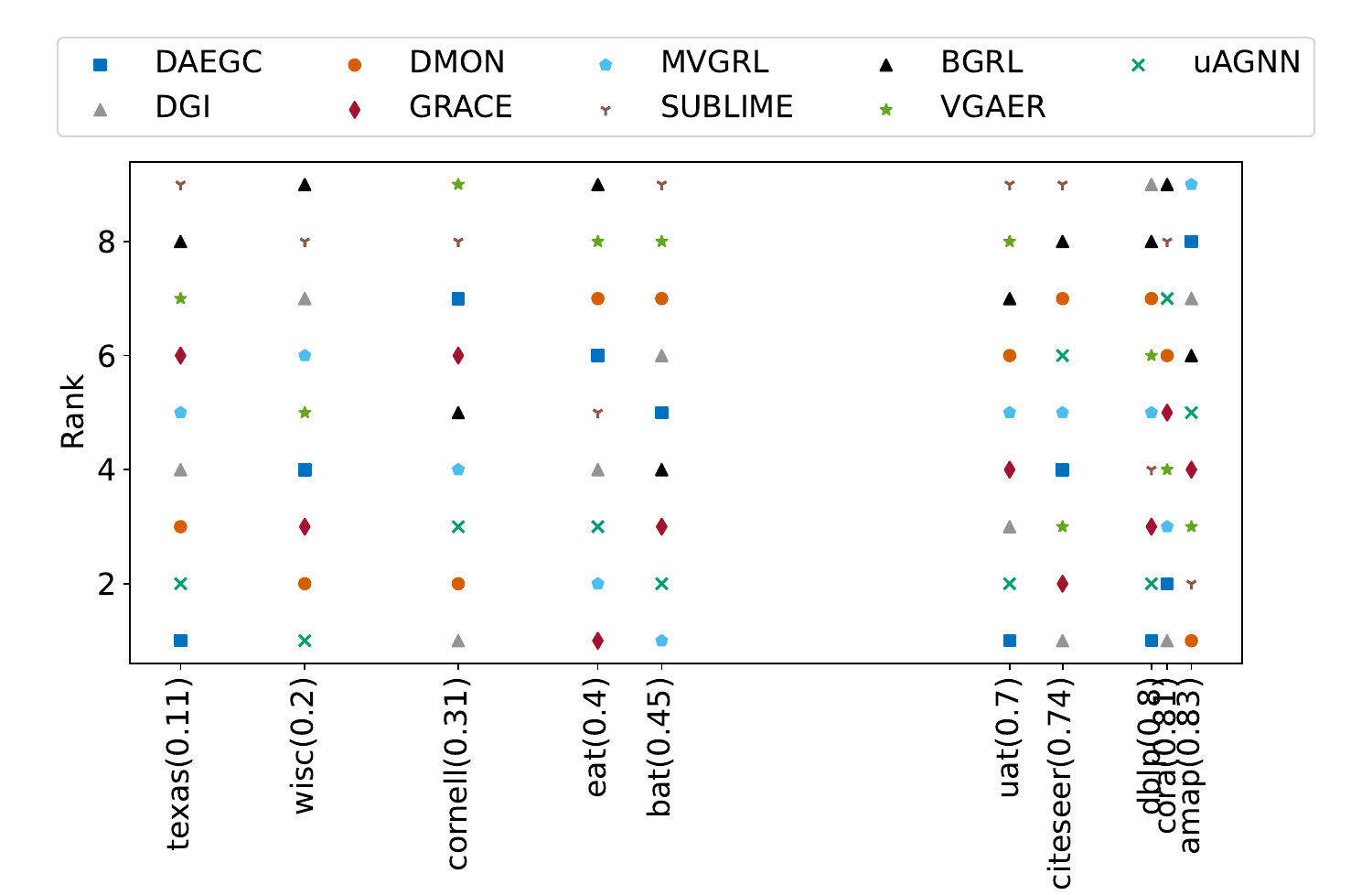}
    \vspace{-4mm}
    \caption{\small F1 score}
    \label{fig:sub1_rank}
  \end{subfigure}
\end{adjustbox}
  
\begin{adjustbox}{center}
   \begin{subfigure}{1.34\columnwidth}
    \centering \includegraphics[width=0.8\linewidth]{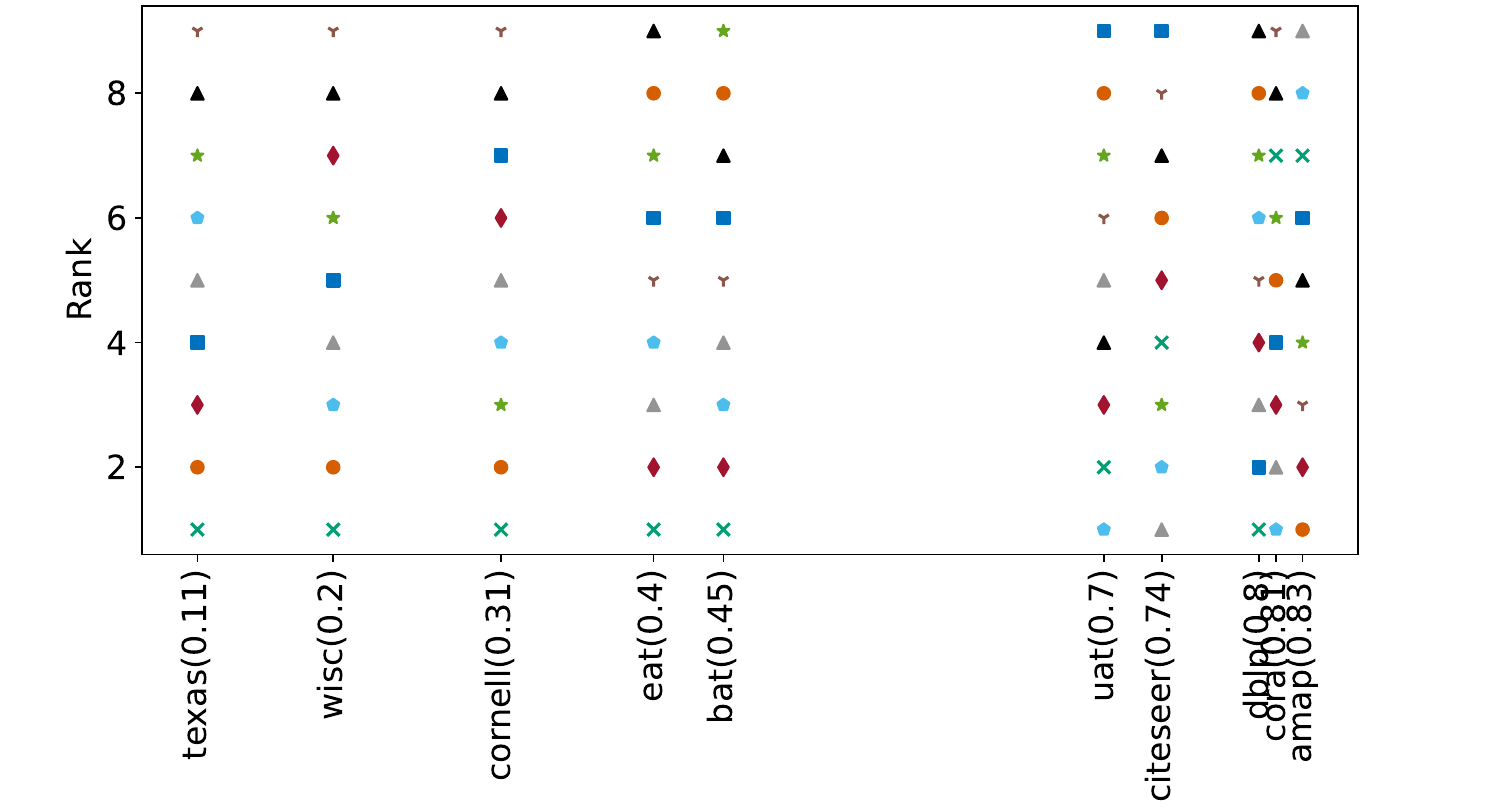}
    \vspace{-4mm}
    \caption{\small NMI score}
    \label{fig:sub2_rank}
  \end{subfigure}
\end{adjustbox}
 
\begin{adjustbox}{center}
   \begin{subfigure}{1.34\columnwidth}
    \centering \includegraphics[width=0.8\linewidth]{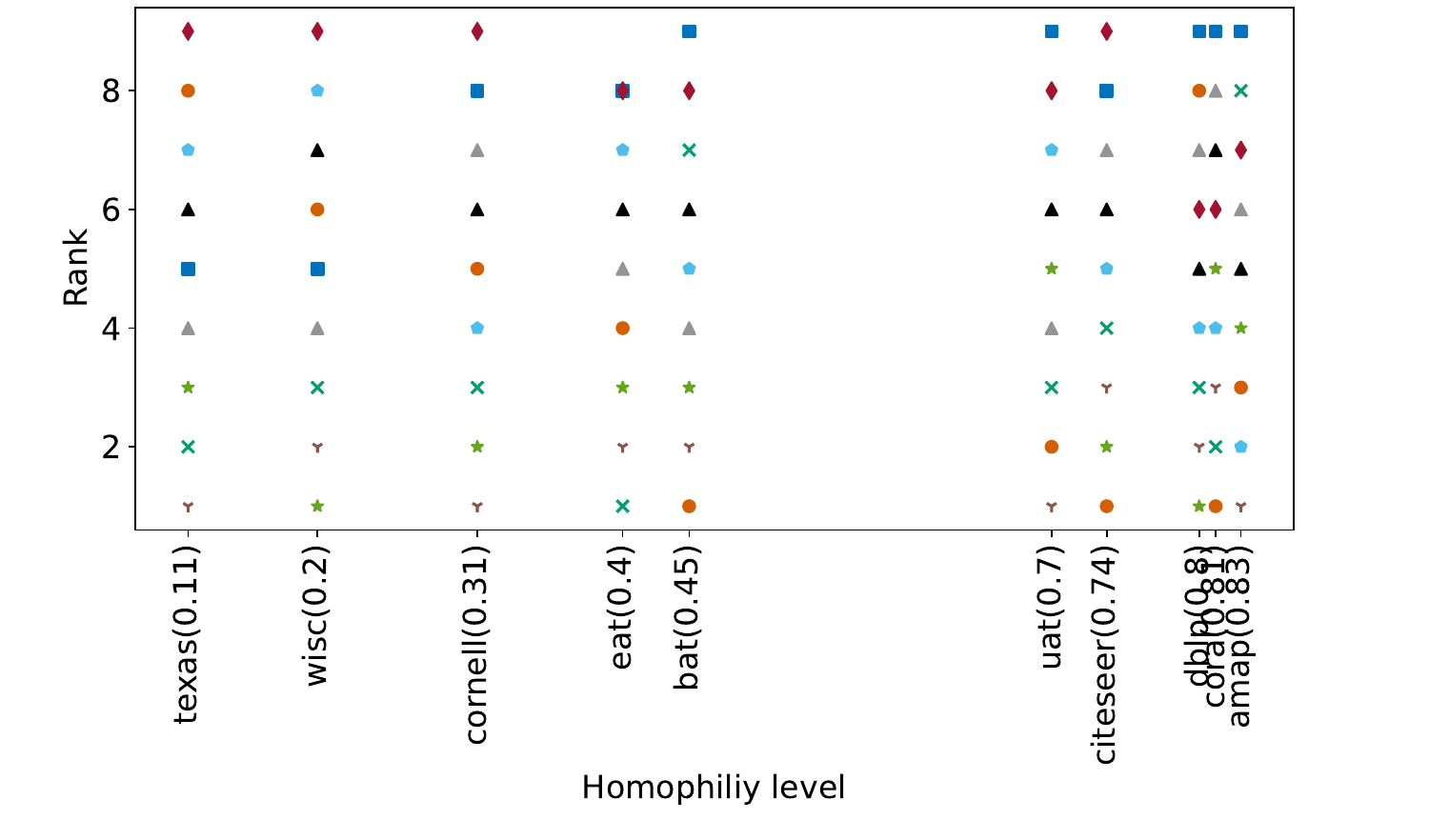}    
    \vspace{-4mm}
    \caption{\small Conductance score}
    \label{fig:sub3_rank}
  \end{subfigure}
\end{adjustbox}
\caption{The rank of the performance of all evaluated methods on the ten considered datesets. The lower, the better.
\label{fig:rank}
}
\end{figure}

\begin{figure}[htpb]
    \centering
\begin{adjustbox}{center}
  \begin{subfigure}{1.34\columnwidth}
    \centering \includegraphics[width=0.8\linewidth]{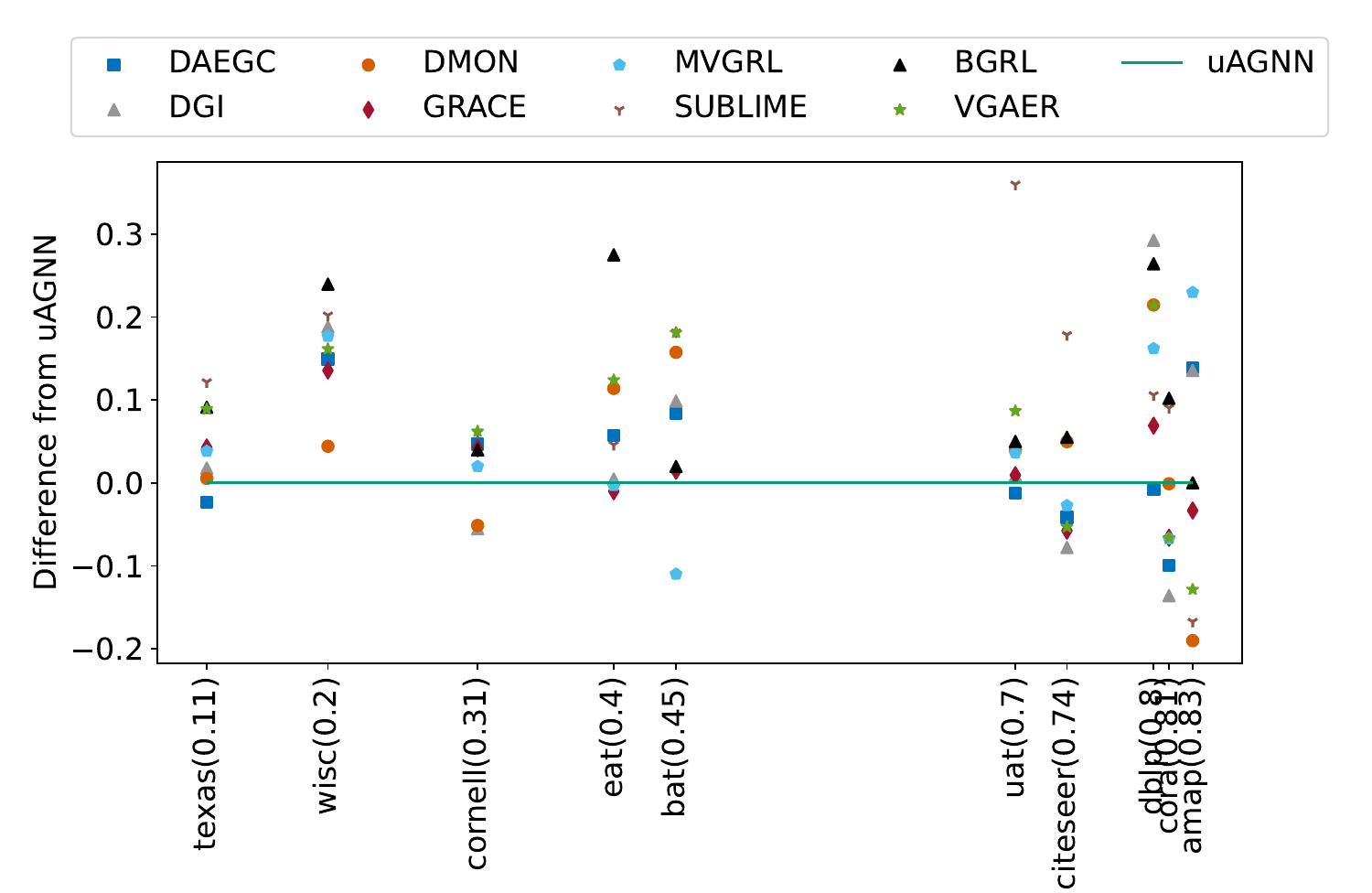}
    \vspace{-4mm}
    \caption{\small F1 score}
    \label{fig:sub1_diff}
  \end{subfigure}
\end{adjustbox}

\begin{adjustbox}{center}
  \begin{subfigure}{1.34\columnwidth}
    \centering \includegraphics[width=0.8\linewidth]{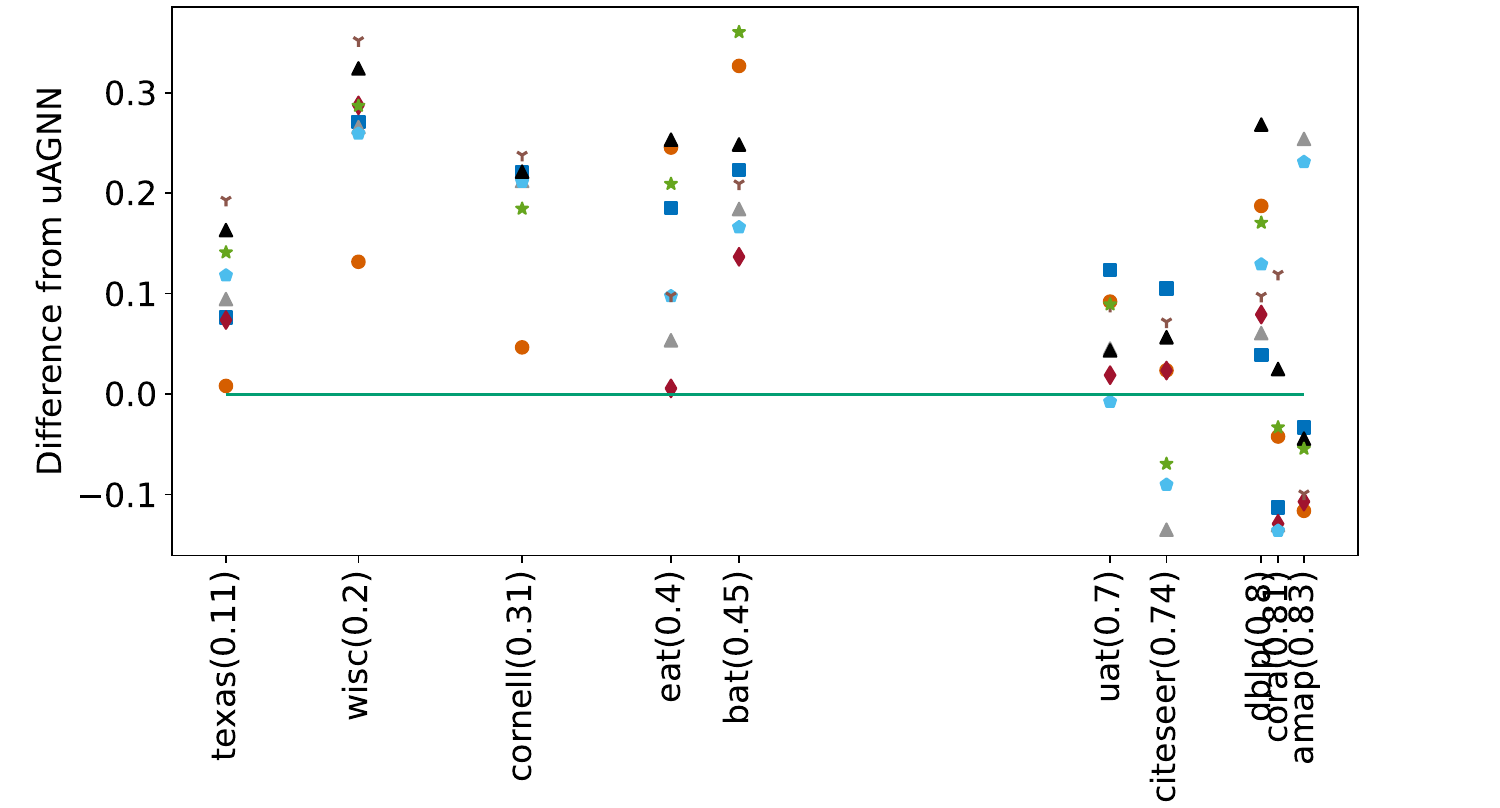}
    \vspace{-4mm}
    \caption{\small NMI score}
    \label{fig:sub2_diff}
  \end{subfigure}
\end{adjustbox}

\begin{adjustbox}{center}
  \begin{subfigure}{1.34\columnwidth}    
  \centering \includegraphics[width=0.8\linewidth]{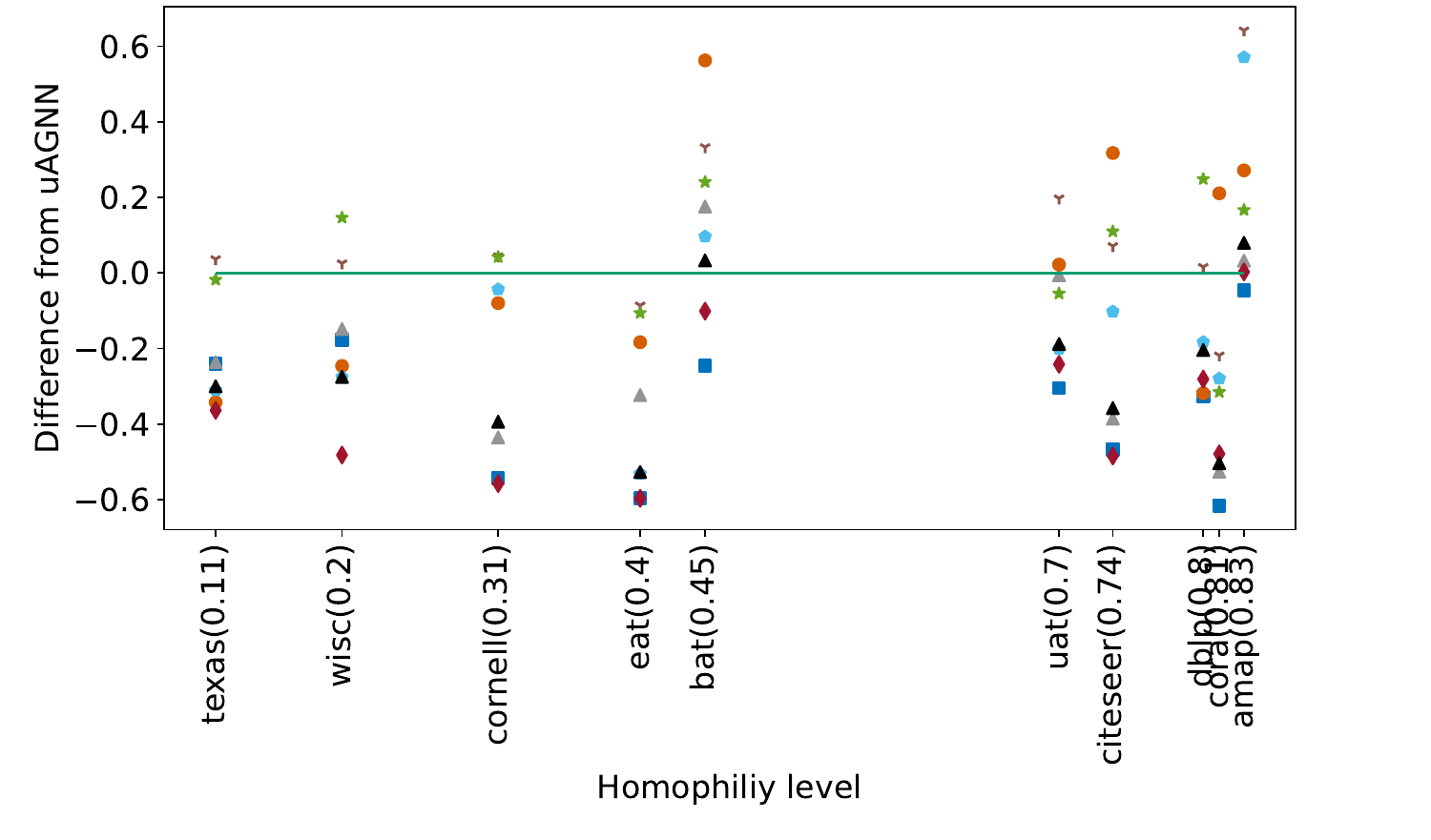}
  \vspace{-4mm}
  \caption{\small Conductance score}
    \label{fig:sub3_diff}
  \end{subfigure}
\end{adjustbox}
\caption{The difference in performance between our uAGNN and literature methods on the ten considered datesets. For F1 and NMI scores, more values above the line indicate better uAGNN results, vice versa for Conductance.
\label{fig:difference}
}
\vspace{-5mm}
\end{figure}

We present the results of our uAGNN with respect to baselines from the literature on the ten different datasets in Table~\ref{tab:f1} (reporting the macro-F1 score), Table~\ref{tab:nmi} (reporting the NMI score), and Table~\ref{tab:conductance} (reporting the conductance score). Additionally, Figure~\ref{fig:rank} illustrates the ranking of all evaluated methods, while Figure~\ref{fig:difference} highlights the performance differences between our method and the baselines.


We observe that when considering the performance with respect to the F1 score, our uAGNN consistently performs favorably across datasets, especially those with low homophily levels, such as Texas ($\mathcal{H}$ = 0.11) and Wisconsin ($\mathcal{H}$ = 0.20) datasets where uAGNN achieves the second-best and best scores, respectively. This trend is further evidenced in both Figures \ref{fig:sub1_rank} and \ref{fig:sub1_diff}, highlighting uAGNN's ability to effectively capture long-range dependencies in graphs, which is critical for low-homophily scenarios.  This phenomenon can be attributed to the structural characteristics of heterophilic graphs, where nodes within the same community are frequently not densely connected. Instead, these nodes are linked through longer, sparser paths. Conversely, uAGNN's performance drops slightly as the homophily level drastically increase, such as in AMAP ($\mathcal{H}$ = 0.83), indicating that models tailored to local neighborhood interactions can maintain an advantage in such scenarios.

The NMI scores further corroborate these findings. uAGNN excels in low to medium homophily levels, \ie up to $\mathcal{H}$ = 0.70, where it effectively balances capturing local and global graph structure. Similar to the F1 score trends, on high-homophily datasets, \ie $\mathcal{H}\geq0.70$, the model's NMI score remains competitive but not dominant. This, again, highlights that the primary strength of our method lies in addressing heterophilous networks, where leveraging long-range information is crucial for achieving superior performance.

Even when considering the conductance as a performance metric, which measures the quality of community separation, our results demonstrate the ability of our method to leverage long-range information to form coherent communities. This is achieved by effectively reducing the number of edges crossing communities, especially when neighboring nodes belong to different classes and distant nodes are part of the same community.

Taken together, these results underline uAGNN's strengths in tackling heterophilous settings. We believe that the key challenge in this setting lies in identifying communities that are not localized and often defined by sparse or long-distance connections. This arises from the fact that the sparsity of intra-community connections and the presence of edges between nodes of different communities can lead to misclassification of community members and reduced precision when using models with limited information propagation capabilities. On the contrary, the exploitation of long-range information (achieved by our uAGNN) helps by incorporating signals from distant but structurally or semantically relevant nodes, allowing the model to pierce together the true community structure. 

\begin{table*}[h!] 
    \centering
    \caption{The mean test F1 score and std averaged over 10 different experimental seeds. The higher, the better.}
   \setlength{\tabcolsep}{2pt}
    \begin{tabular}{ |c | cccccccccc | }
    \hline
\multirow{2}{*}{\textbf{Model}} & \textbf{Texas} & \textbf{Wisconsin} &\textbf{Cornell} & \textbf{EAT} & \textbf{BAT} & \textbf{UAT} & \textbf{CiteSeer} & \textbf{DBLP} & \textbf{Cora} & \textbf{AMAP}  \\
 & {\scriptsize$\mathcal{H}$ = 0.11} & {\scriptsize$\mathcal{H}$ = 0.20} &  {\scriptsize$\mathcal{H}$ = 0.31} &  {\scriptsize$\mathcal{H}$ = 0.40} &  {\scriptsize$\mathcal{H}$ = 0.45} &  {\scriptsize$\mathcal{H}$ = 0.70} &  {\scriptsize$\mathcal{H}$ = 0.74} &  {\scriptsize$\mathcal{H}$ = 0.80} &  {\scriptsize$\mathcal{H}$ = 0.81} & {\scriptsize$\mathcal{H}$ = 0.83}\\\hline
 DAEGC & \textbf{0.35$_{\pm0.05}$} & 0.28$_{\pm0.02}$ & 0.24$_{\pm0.03}$ & 0.41$_{\pm0.02}$ & 0.46$_{\pm0.04}$ & \textbf{0.54$_{\pm0.01}$} & 0.60$_{\pm0.02}$ & \textbf{0.72$_{\pm0.05}$} & 0.58$_{\pm0.08}$ & 0.34$_{\pm0.02}$ \\
DGI & 0.31$_{\pm0.03}$ & 0.24$_{\pm0.03}$ & \textbf{0.34$_{\pm0.06}$} & 0.46$_{\pm0.05}$ & 0.44$_{\pm0.06}$ & 0.52$_{\pm0.03}$ & \textbf{0.63$_{\pm0.01}$} & 0.42$_{\pm0.21}$ & \textbf{0.61$_{\pm0.04}$} & 0.34$_{\pm0.06}$ \\
DMON & 0.33$_{\pm0.02}$ & 0.39$_{\pm0.03}$ & \textbf{0.34$_{\pm0.05}$} & 0.35$_{\pm0.04}$ & 0.38$_{\pm0.07}$ & 0.49$_{\pm0.04}$ & 0.50$_{\pm0.07}$ & 0.50$_{\pm0.07}$ & 0.48$_{\pm0.06}$ & \textbf{0.67$_{\pm0.10}$} \\
GRACE & 0.29$_{\pm0.08}$ & 0.29$_{\pm0.01}$ & 0.25$_{\pm0.03}$ & \textbf{0.48$_{\pm0.03}$} & 0.53$_{\pm0.03}$ & 0.52$_{\pm0.03}$ & 0.61$_{\pm0.03}$ & 0.64$_{\pm0.07}$ & 0.54$_{\pm0.03}$ & 0.51$_{\pm0.09}$ \\
MVGRL & 0.29$_{\pm0.02}$ & 0.25$_{\pm0.05}$ & 0.27$_{\pm0.05}$ & 0.47$_{\pm0.06}$ & \textbf{0.65$_{\pm0.05}$} & 0.49$_{\pm0.04}$ & 0.58$_{\pm0.02}$ & 0.55$_{\pm0.07}$ & 0.54$_{\pm0.03}$ & 0.25$_{\pm0.03}$ \\
SUBLIME & 0.21$_{\pm0.03}$ & 0.23$_{\pm0.02}$ & 0.23$_{\pm0.03}$ & 0.42$_{\pm0.06}$ & 0.36$_{\pm0.06}$ & 0.17$_{\pm0.13}$ & 0.38$_{\pm0.15}$ & 0.60$_{\pm0.11}$ & 0.39$_{\pm0.05}$ & 0.65$_{\pm0.07}$ \\
BGRL & 0.24$_{\pm0.03}$ & 0.19$_{\pm0.06}$ & 0.25$_{\pm0.02}$ & 0.19$_{\pm0.02}$ & 0.52$_{\pm0.03}$ & 0.48$_{\pm0.02}$ & 0.50$_{\pm0.04}$ & 0.45$_{\pm0.06}$ & 0.38$_{\pm0.06}$ & 0.48$_{\pm0.10}$ \\
VGAER & 0.24$_{\pm0.03}$ & 0.27$_{\pm0.03}$ & 0.22$_{\pm0.02}$ & 0.34$_{\pm0.03}$ & 0.36$_{\pm0.06}$ & 0.44$_{\pm0.03}$ & 0.61$_{\pm0.14}$ & 0.50$_{\pm0.13}$ & 0.54$_{\pm0.14}$ & 0.61$_{\pm0.16}$ \\\hline
uAGNN & 0.33$_{\pm0.09}$ & \textbf{0.43$_{\pm0.06}$} & 0.29$_{\pm0.04}$ & 0.47$_{\pm0.05}$ & 0.54$_{\pm0.10}$ & 0.53$_{\pm0.02}$ & 0.55$_{\pm0.11}$ & 0.71$_{\pm0.14}$ & 0.48$_{\pm0.08}$ & 0.48$_{\pm0.09}$ \\
\hline
\end{tabular}
    \label{tab:f1}
\end{table*}

\begin{table*}[h!]
    \centering
   \setlength{\tabcolsep}{2pt}
    \caption{The mean test NMI score and std averaged over 10 different experimental seeds. The higher, the better.}
    \begin{tabular}{ |c | cccccccccc | }
    \hline
\multirow{2}{*}{\textbf{Model}} & \textbf{Texas} & \textbf{Wisconsin} &\textbf{Cornell} & \textbf{EAT} & \textbf{BAT} & \textbf{UAT} & \textbf{CiteSeer} & \textbf{DBLP} & \textbf{Cora} & \textbf{AMAP}  \\
 & {\scriptsize$\mathcal{H}$ = 0.11} & {\scriptsize$\mathcal{H}$ = 0.20} &  {\scriptsize$\mathcal{H}$ = 0.31} &  {\scriptsize$\mathcal{H}$ = 0.40} &  {\scriptsize$\mathcal{H}$ = 0.45} &  {\scriptsize$\mathcal{H}$ = 0.70} &  {\scriptsize$\mathcal{H}$ = 0.74} &  {\scriptsize$\mathcal{H}$ = 0.80} &  {\scriptsize$\mathcal{H}$ = 0.81} & {\scriptsize$\mathcal{H}$ = 0.83}\\\hline
DAEGC & 0.14$_{\pm0.02}$ & 0.11$_{\pm0.02}$ & 0.05$_{\pm0.02}$ & 0.13$_{\pm0.02}$ & 0.25$_{\pm0.04}$ & 0.16$_{\pm0.04}$ & 0.19$_{\pm0.04}$ & 0.37$_{\pm0.05}$ & 0.47$_{\pm0.02}$ & 0.53$_{\pm0.02}$ \\
DGI & 0.12$_{\pm0.02}$ & 0.11$_{\pm0.04}$ & 0.06$_{\pm0.01}$ & 0.26$_{\pm0.02}$ & 0.29$_{\pm0.05}$ & 0.24$_{\pm0.03}$ & \textbf{0.43$_{\pm0.01}$} & 0.35$_{\pm0.18}$ & 0.49$_{\pm0.01}$ & 0.24$_{\pm0.08}$ \\
DMON & 0.21$_{\pm0.04}$ & 0.24$_{\pm0.05}$ & 0.22$_{\pm0.07}$ & 0.07$_{\pm0.02}$ & 0.15$_{\pm0.05}$ & 0.20$_{\pm0.04}$ & 0.27$_{\pm0.03}$ & 0.22$_{\pm0.09}$ & 0.40$_{\pm0.04}$ &\textbf{ 0.61$_{\pm0.08}$} \\
GRACE & 0.15$_{\pm0.00}$ & 0.09$_{\pm0.02}$ & 0.05$_{\pm0.01}$ & 0.31$_{\pm0.02}$ & 0.34$_{\pm0.06}$ & 0.27$_{\pm0.01}$ & 0.27$_{\pm0.10}$ & 0.33$_{\pm0.05}$ & 0.49$_{\pm0.04}$ & 0.60$_{\pm0.04}$ \\
MVGRL & 0.10$_{\pm0.02}$ & 0.12$_{\pm0.03}$ & 0.06$_{\pm0.02}$ & 0.22$_{\pm0.06}$ & 0.31$_{\pm0.07}$ & \textbf{0.29$_{\pm0.02}$} & 0.39$_{\pm0.02}$ & 0.28$_{\pm0.03}$ & \textbf{0.50$_{\pm0.03}$} & 0.27$_{\pm0.08}$ \\
SUBLIME & 0.03$_{\pm0.01}$ & 0.02$_{\pm0.01}$ & 0.03$_{\pm0.01}$ & 0.22$_{\pm0.02}$ & 0.27$_{\pm0.05}$ & 0.20$_{\pm0.05}$ & 0.23$_{\pm0.14}$ & 0.31$_{\pm0.02}$ & 0.24$_{\pm0.05}$ & 0.60$_{\pm0.04}$ \\
BGRL & 0.06$_{\pm0.03}$ & 0.05$_{\pm0.03}$ & 0.05$_{\pm0.01}$ & 0.06$_{\pm0.02}$ & 0.23$_{\pm0.05}$ & 0.24$_{\pm0.00}$ & 0.24$_{\pm0.02}$ & 0.14$_{\pm0.02}$ & 0.34$_{\pm0.06}$ & 0.54$_{\pm0.02}$ \\
VGAER & 0.08$_{\pm0.02}$ & 0.09$_{\pm0.04}$ & 0.08$_{\pm0.03}$ & 0.11$_{\pm0.06}$ & 0.12$_{\pm0.06}$ & 0.20$_{\pm0.02}$ & 0.37$_{\pm0.13}$ & 0.24$_{\pm0.11}$ & 0.39$_{\pm0.14}$ & 0.55$_{\pm0.18}$ \\\hline
uAGNN & \textbf{0.22$_{\pm0.07}$} & \textbf{0.38$_{\pm0.07}$} & \textbf{0.27$_{\pm0.07}$} & \textbf{0.32$_{\pm0.01}$} & \textbf{0.48$_{\pm0.12}$} & \textbf{0.29$_{\pm0.02}$} & 0.30$_{\pm0.08}$ & \textbf{0.41$_{\pm0.12}$} & 0.36$_{\pm0.10}$ & 0.50$_{\pm0.10}$ \\
\hline
\end{tabular}
    \label{tab:nmi}
\end{table*}

\begin{table*}[h!]
    \centering
    \setlength{\tabcolsep}{2pt}
    \caption{The mean test conductance score and std averaged over 10 different experimental seeds. The lower, the better.}
    \begin{tabular}{ |c | cccccccccc | }
    \hline
    \multirow{2}{*}{\textbf{Model}} & \textbf{Texas} & \textbf{Wisconsin} &\textbf{Cornell} & \textbf{EAT} & \textbf{BAT} & \textbf{UAT} & \textbf{CiteSeer} & \textbf{DBLP} & \textbf{Cora} & \textbf{AMAP}  \\
 & {\scriptsize$\mathcal{H}$ = 0.11} & {\scriptsize$\mathcal{H}$ = 0.20} &  {\scriptsize$\mathcal{H}$ = 0.31} &  {\scriptsize$\mathcal{H}$ = 0.40} &  {\scriptsize$\mathcal{H}$ = 0.45} &  {\scriptsize$\mathcal{H}$ = 0.70} &  {\scriptsize$\mathcal{H}$ = 0.74} &  {\scriptsize$\mathcal{H}$ = 0.80} &  {\scriptsize$\mathcal{H}$ = 0.81} & {\scriptsize$\mathcal{H}$ = 0.83}\\\hline
DAEGC & 0.70$_{\pm0.15}$ & 0.70$_{\pm0.06}$ & 0.99$_{\pm0.02}$ & 1.00$_{\pm0.00}$ & 1.00$_{\pm0.00}$ & 1.00$_{\pm0.00}$ & 0.98$_{\pm0.04}$ & 1.00$_{\pm0.00}$ & 1.00$_{\pm0.00}$ & 0.92$_{\pm0.14}$ \\
DGI & 0.70$_{\pm0.29}$ & 0.67$_{\pm0.12}$ & 0.88$_{\pm0.23}$ & 0.73$_{\pm0.22}$ & 0.58$_{\pm0.14}$ & 0.70$_{\pm0.15}$ & 0.90$_{\pm0.05}$ & 0.98$_{\pm0.01}$ & 0.91$_{\pm0.16}$ & 0.84$_{\pm0.26}$ \\
DMON & 0.80$_{\pm0.21}$ & 0.76$_{\pm0.19}$ & 0.52$_{\pm0.37}$ & 0.59$_{\pm0.48}$ & \textbf{0.19$_{\pm0.38}$} & 0.67$_{\pm0.44}$ & \textbf{0.20$_{\pm0.40}$} & 0.99$_{\pm0.02}$ & \textbf{0.17$_{\pm0.35}$} & 0.60$_{\pm0.49}$ \\
GRACE & 0.82$_{\pm0.30}$ & 1.00$_{\pm0.00}$ & 1.00$_{\pm0.00}$ & 1.00$_{\pm0.00}$ & 0.86$_{\pm0.24}$ & 0.94$_{\pm0.19}$ & 1.00$_{\pm0.00}$ & 0.95$_{\pm0.03}$ & 0.86$_{\pm0.28}$ & 0.87$_{\pm0.13}$ \\
MVGRL & 0.77$_{\pm0.24}$ & 0.79$_{\pm0.30}$ & 0.49$_{\pm0.34}$ & 0.93$_{\pm0.16}$ & 0.66$_{\pm0.13}$ & 0.90$_{\pm0.03}$ & 0.62$_{\pm0.20}$ & 0.86$_{\pm0.11}$ & 0.66$_{\pm0.04}$ & 0.30$_{\pm0.02}$ \\
SUBLIME & \textbf{0.42$_{\pm0.25}$} & 0.49$_{\pm0.21}$ & \textbf{0.40$_{\pm0.32}$} & 0.49$_{\pm0.19}$ & 0.42$_{\pm0.13}$ & \textbf{0.50$_{\pm0.20}$ }& 0.44$_{\pm0.28}$ & 0.66$_{\pm0.18}$ & 0.60$_{\pm0.39}$ & \textbf{0.23$_{\pm0.18}$} \\
BGRL & 0.76$_{\pm0.11}$ & 0.79$_{\pm0.16}$ & 0.84$_{\pm0.14}$ & 0.93$_{\pm0.01}$ & 0.72$_{\pm0.15}$ & 0.88$_{\pm0.11}$ & 0.87$_{\pm0.04}$ & 0.88$_{\pm0.02}$ & 0.89$_{\pm0.02}$ & 0.79$_{\pm0.06}$ \\
VGAER & 0.48$_{\pm0.10}$ & \textbf{0.37$_{\pm0.17}$} & \textbf{0.40$_{\pm0.10}$} & 0.51$_{\pm0.20}$ & 0.51$_{\pm0.15}$ & 0.75$_{\pm0.12}$ & 0.41$_{\pm0.20}$ & \textbf{0.42$_{\pm0.08}$} & 0.70$_{\pm0.19}$ & 0.70$_{\pm0.18}$ \\\hline
uAGNN & 0.46$_{\pm0.14}$ & 0.52$_{\pm0.07}$ & 0.44$_{\pm0.15}$ & \textbf{0.40$_{\pm0.37}$} & 0.75$_{\pm0.17}$ & 0.69$_{\pm0.23}$ & 0.52$_{\pm0.15}$ & 0.67$_{\pm0.14}$ & 0.38$_{\pm0.20}$ & 0.87$_{\pm0.08}$ \\
\hline
\end{tabular}
    \label{tab:conductance}
\end{table*}

\section{Conclusions}
This work explored the role of non-local node interactions in community detection tasks when tackled by message passing GNNs. We highlighted the importance of being able to effectively propagate information between nodes across longer paths when attempting the unsupervised identification of such sparsely connected components, highlighting the similarity with the broadly studied heterophilic setting in supervised node classification. Folloing up these considerations, we introduced the \textit{Unsupervised Antisymmetric Graph Neural Network} (uAGNN), a novel framework for community detection leveraging non-dissipative GNNs. Unlike prior approaches in the community detection domain, our framework imposes stability and conservative constraints on the GNN through antisymmetric weight matrices, enabling it to learn and retain long-range dependencies between nodes effectively.
Our experimental analysis demonstrates that the stable, non-dissipative architecture of uAGNN delivers superior performance in low-homophily settings. Overall, uAGNN emerges as a highly effective framework for unsupervised community detection, particularly in scenarios where traditional methods struggle to capture long-range dependencies. These findings underscore its potential for advancing graph representation learning and tackling challenges inherent to diverse graph structures, such as heterophilic graphs.

Looking ahead to future developments, we aim to extend this framework by exploring non-dissipative architectures derived from alternative discretization methods, such as multistep schemes \cite{Ascher1998}. Additionally, we plan to investigate novel backbone architectures that can balance information conservation and dissipation during propagation, as in \cite{gravina_phdgn}, to enhance performance in settings dominated by local neighborhood interactions.

\section*{Acknowledgment}
The work has been partially supported by the EU project EMERGE (Grant No. 101070918).

\bibliographystyle{ieeetr}
\bibliography{references}

\end{document}